\documentclass[10pt, conference, letterpaper]{IEEEtran}
\usepackage{cite}
\usepackage{amsmath,amssymb,bm}
\usepackage{graphicx}
\usepackage{booktabs,array}
\usepackage[table]{xcolor}
\usepackage{tikz}
\usetikzlibrary{arrows.meta,patterns}
\usepackage[hidelinks]{hyperref}
\usepackage{orcidlink}
\hypersetup{pdftitle={Physical Sensing Verification for Cognitive ISAC against UAV Spoofing via Digital Twins},pdfauthor={},pdfsubject={},pdfkeywords={},pdfcreator={},pdfproducer={}}

\definecolor{pendingcolor}{RGB}{145,66,17}
\definecolor{tabhead}{RGB}{208,226,243}
\definecolor{tabrow}{RGB}{238,245,252}
\definecolor{targetblue}{RGB}{0,114,178}
\definecolor{relayred}{RGB}{213,94,0}

\newcommand{\HT}{\mathcal H_{\mathrm T}}
\newcommand{\HA}{\mathcal H_{\mathrm A}}
\newcommand{\DT}{\widehat{\mathcal E}}
\newcommand{\norm}[1]{\left\lVert#1\right\rVert}
\newcommand{\pB}{\mathbf p_{\rm B}}
\newcommand{\pT}{\mathbf p_{\rm T}}
\newcommand{\pA}{\mathbf p_{\rm A}}
\newtheorem{proposition}{Proposition}

\begin{document}

\title{Physical Sensing Verification for Cognitive ISAC\\against UAV Spoofing via Digital Twins}

\author{
\IEEEauthorblockN{Seyed Bagher Hashemi Natanzi\,\orcidlink{0000-0003-1524-8669}, Bo Tang\,\orcidlink{0000-0001-5708-766X}}
\IEEEauthorblockA{Department of Electrical and Computer Engineering, Worcester Polytechnic Institute (WPI), USA\\
Email: \{snatanzi, btang1\}@wpi.edu}
}

\maketitle
\maketitle
    \begin{tikzpicture}[remember picture, overlay]
        \node[anchor=north, yshift=-0.5cm] at (current page.north) {\fbox{\parbox{\textwidth}{\centering\small {\color{red}This work has been submitted to the IEEE Conference for possible publication. Copyright may be transferred without notice, after which this version may no longer be accessible.}}}};
    \end{tikzpicture}
\begin{abstract}
As integrated sensing and communication (ISAC) lets 5G and 6G base stations detect drones, software agents will act on these detections, so a replayed signal can plant a phantom drone that triggers real actions. This paper proposes a wall-based physical check for such cognitive ISAC systems. A digital twin (DT) predicts the delay and angle of a genuine target's wall echo. A relay can fake the direct echo's range but, standing elsewhere, not the direction of its own wall echo. The paper proves that line-of-sight beam steering cannot expose a matched relay, and that, under stated conditions, a single-antenna relay faking a wall echo leaves a telltale echo where a genuine target leaves none. In a pre-registered ray-tracing study, the twin's evidence separated genuine targets and replays perfectly (AUC 1.000), and a later-registered passive check without extra transmission accepted every held-out genuine target and no replay. The primary active probe was inconclusive (5.3\% false acceptances against a 5\% limit), lacking threshold margin. At the evaluated relay positions, the conditional bound certifies 3 of 24 held-out streets against adaptive or fixed single-antenna relay signals meeting its conditions, reproduced on fresh streets. Doubling the bandwidth raises this to 11 of 24.
\end{abstract}


\section{Introduction}

Integrated sensing and communication (ISAC) turns a cellular base station into a radar: from the echoes of its own orthogonal frequency-division multiplexing (OFDM) signals, it detects drones that carry no transponder~\cite{tr22837,liu2022isac}. Vendors and operators have already shown live drone detection and tracking on 5G massive multiple-input multiple-output (MIMO) base stations, with artificial intelligence (AI) classification at the network edge~\cite{fierce2026}. They have also shown how such sensing could feed the command and control systems of public authorities~\cite{ericsson2026}. The next step is to let software act on these detections. An ISAC system is \emph{cognitive} when it closes the loop of sensing, predicting, and acting, as cognitive radar does~\cite{haykin2006}. In such a system, an autonomous controller turns sensed objects into radio actions such as beam and tracking assignments, whether it follows fixed rules or is an AI agent built on a large language model (LLM).

Autonomy makes fake detections dangerous, and they are cheap. A digital radio frequency memory (DRFM) relay captures the transmitted waveform, delays it, and sends it back, so the base station sees an echo at a range of the attacker's choosing. Such replay attacks on OFDM sensing have been shown in simulation and over the air~\cite{chrysanidis2024,bartoletti2026,yildirim2025}. If the relay sits on the line toward the claimed position and matches the delay and strength of a real echo, the direct echo looks exactly right. A controller that trusts it may waste beams on a phantom or raise a false alarm. The problem is therefore: before acting, how can a single base station check, from physical evidence alone, that something is really at the claimed position?

Existing defenses use other clues: residual carrier frequency offset (CFO) left by the relay hardware~\cite{argyriou2023}, trajectory consistency against spoofing by a reconfigurable intelligent surface (RIS)~\cite{shui2025}, or active RIS changes against forged channel state information (CSI)~\cite{liu2026}; radar adds waveform agility and multi-sensor checks~\cite{calatrava2025}. Closest to this paper, multistatic radars reject false targets because a single jammer cannot reproduce how a real target scatters toward receivers at different places~\cite{zhao2015,zhao2016}. Multistatic cooperation is also the first countermeasure proposed against blind ISAC replay~\cite{bartoletti2026}. These checks, however, rely on hardware imperfections, extra hardware, time to build a track, or several receiving sites.

This paper uses a resource that every urban base station already has: the walls around it. Operators report that reflections off buildings degrade drone tracking in dense cities~\cite{delooper2026att}; this paper takes the opposite view and treats walls as mirrors. A digital twin (DT), a geometric and material model of the nearby walls that may contain errors~\cite{alkhateeb2023dt}, predicts the delay and angle at which a genuine target's wall echo must arrive. A relay can fake range by adding delay, but its own wall echo comes from where it actually stands. The walls thus act as virtual receivers, much as virtual anchors do in multipath localization~\cite{leitinger2019}, and give one base station the spatial check that multistatic radar gets from several sites. The aim of this paper is to turn this idea into a verifier that a controller can call before it acts, and to test where it holds and where it fails. The contributions are:
\begin{enumerate}
\item two results on what beams can reveal: steering along the line of sight (LoS) cannot expose a matched relay, and a relay's single-wall echo always arrives later than a genuine target's;
\item a twin-guided check that reads the predicted wall echo from the echo already received, with an optional extra probe through the wall;
\item a coupling bound showing that, under stated conditions, a single-antenna relay with any filter must leave a telltale echo to fake the wall echo, and a conservative certificate built on it, with a measurement of where it currently holds;
\item a pre-registered evaluation with physically propagated relay signals that shows where this evidence holds and where it fails.
\end{enumerate}
The paper builds not the controller but the evidence that any controller, rule-based or AI-driven, should request before acting. Propagation is simulated with Sionna RT~2.2~\cite{hoydis2023}; code and the analysis plan are public.\footnote{\url{https://github.com/CLIS-WPI/isac-wall-pilot}}

\section{System and Threat Model}
\label{sec:model}

A monostatic base station (gNB, the 5G New Radio base station; the name for 6G is not yet fixed) at $\pB$ has co-located transmit and receive arrays with responses $\mathbf a_{\rm T}(\phi)\in\mathbb C^{N_{\rm t}}$ and $\mathbf a_{\rm R}(\theta)\in\mathbb C^{N_{\rm r}}$, where $\phi$ is a departure angle and $\theta$ an arrival angle; with $\mathbf v(\phi)=\mathbf a_{\rm T}^*(\phi)$, a beam $\mathbf w$ has power gain $|\mathbf v^{\rm H}(\phi)\mathbf w|^2$. An \emph{action} transmits $M$ OFDM symbols over $N$ subcarriers of spacing $\Delta f$ (bandwidth $B=N\Delta f$, baseband frequencies $f_n$), with fresh known unit-modulus symbols $s[n,m]$ and a beam $\mathbf w_b$ of power $\norm{\mathbf w_b}^2=P_{\rm tx}$. Receiver noise has power $k_{\rm B}T_0BF$, with Boltzmann constant $k_{\rm B}$, temperature $T_0$, and noise figure $F$.

The environment $\mathcal E$ contains finite planar reflectors; reflector $k$ has unit normal $\mathbf n_k$ and contains the point $\mathbf q_k$. The twin $\DT$ shifts each reflector along its normal by an independent offset $\epsilon_k\sim\mathcal N(0,\sigma_{\rm map}^2)$. The verifier knows $\sigma_{\rm map}$ but not the offsets.

Under the target hypothesis $\HT$, a point scatterer with radar cross-section (RCS) $\sigma_{\rm T}$ sits at $\pT=\pB+R_{\rm T}\mathbf u$. Under the attack hypothesis $\HA$, a relay sits on the same bearing $\mathbf u$ but closer, at $\pA=\pB+R_{\rm A}\mathbf u$ with $R_{\rm A}<R_{\rm T}$, so that its direct echo arrives from the right angle. Two relay classes are considered. A \emph{fixed-filter} relay applies one gain $g$ and one delay $\delta\ge\tau_{\rm proc}$ throughout, with output power below $P_{\rm A}$:
\begin{equation}
\begin{aligned}
\mathbf y_{\rm A}[n,m]={}&\mathbf h_{\rm U}[n,m]\,g e^{-j2\pi f_n\delta}\\
&\times\bigl(\mathbf h_{\rm D}^{\mathsf T}[n,m]\mathbf w_b s[n,m]+z_{\rm A}[n,m]\bigr),
\end{aligned}
\label{eq:relay}
\end{equation}
where $\mathbf h_{\rm D}$ and $\mathbf h_{\rm U}$ are the gNB-relay channels including all multipath and $z_{\rm A}$ is relay noise, which is forwarded; $\delta=2(R_{\rm T}-R_{\rm A})/c$ and a $g$ matched to the LoS path gains make the relay's direct echo equal a real one for every beam, since the relay lies on the target's bearing. A \emph{single-antenna emitter} at $\pA$ may instead transmit any causal signal $x(t)$ and change it at will; with all delays inside the cyclic prefix, it contributes $\mathbf y_{\rm A}[n,m]=\mathbf h_{\rm U}[n,m]x_{\rm A}[n,m]$, where $x_{\rm A}$ holds its subcarrier symbols. Relays with several antennas, or several cooperating relays, are beyond the scope of this paper. The gNB receives
\begin{equation}
\mathbf y[n,m]=(\mathbf H_{\rm T}+\mathbf H_{\rm B})[n,m]\mathbf w_b s[n,m]+\mathbf y_{\rm A}[n,m]+\mathbf z[n,m],
\label{eq:rx}
\end{equation}
where $\mathbf H_{\rm T}$ holds the scatterer's paths (zero under $\HA$), $\mathbf H_{\rm B}$ the static clutter (in the street, specular wall clutter returns from $\pm90^\circ$ within 80 to 120\,ns, outside the surveillance window), and $\mathbf y_{\rm A}=\mathbf 0$ under $\HT$. Each path $\ell$ contributes $\alpha_\ell\mathbf a_{\rm R}(\theta_\ell)\mathbf a_{\rm T}^{\mathsf T}(\phi_\ell)e^{-j2\pi f_n\tau_\ell}$.

\section{Why Walls Help}
\label{sec:physics}

Beams along the line of sight cannot help; walls can.

\begin{proposition}[LoS indistinguishability]
\label{prop:los}
Suppose that under both hypotheses all paths lie along $\mathbf u$, with equal far-field responses, matched delay, Doppler, and complex gain, identical clutter, and equal total noise distributions, including any forwarded relay noise. Then every causal policy that adapts only the transmit beam yields the same observation distribution under $\HT$ and $\HA$.
\end{proposition}

\begin{IEEEproof}
Both channels equal $\beta[n,m]\mathbf a_{\rm R}(\theta_0)\mathbf a_{\rm T}^{\mathsf T}(\phi_0)$, so $\mathbf H\mathbf w_bs$ is the same for every beam; induction over the policy's actions extends the equality to the whole record.
\end{IEEEproof}

Walls break this symmetry. Let $\mathcal M_k(\mathbf x)=\mathbf x-2\mathbf n_k\mathbf n_k^{\mathsf T}(\mathbf x-\mathbf q_k)$ mirror a point across reflector $k$. A \emph{mixed path} to $\mathbf p$ uses reflector $k$ once and the direct line once: the \emph{outgoing} one reflects on the way out, the \emph{reciprocal} one on the way back. The image method replaces a reflection by a straight line to the mirrored point, so both paths have length $\norm{\mathbf p-\mathcal M_k(\pB)}+\norm{\mathbf p-\pB}$ whenever the specular point lies on the reflector, and the reciprocal path arrives from the direction of $\mathcal M_k(\mathbf p)$. With $u_k(\mathbf p)=\norm{\mathbf p-\mathcal M_k(\pB)}-\norm{\mathbf p-\pB}$, the matched fixed-filter relay's mixed-path echo trails the scatterer's by
\begin{equation}
\Delta\tau_k=\bigl[u_k(\pA)-u_k(\pT)\bigr]/c.
\label{eq:gap}
\end{equation}

\begin{proposition}[Delay ordering]
\label{prop:gap}
If the mixed path via reflector $k$ exists for both $\pA$ and $\pT$, the plane of reflector $k$ does not contain $\pB$, and $\mathbf u$ is not parallel to $\mathbf n_k$, then $\Delta\tau_k>0$.
\end{proposition}

\begin{IEEEproof}
Along $\pB+R\mathbf u$, $\mathrm du_k/\mathrm dR=\mathbf u^{\mathsf T}\mathbf e(R)-1\le0$, where $\mathbf e$ is the unit vector from $\mathcal M_k(\pB)$ to $\pB+R\mathbf u$; equality would require $\mathcal M_k(\pB)-\pB$, which is parallel to $\mathbf n_k$, to be parallel to $\mathbf u$, so $u_k$ strictly decreases and $R_{\rm A}<R_{\rm T}$ gives $\Delta\tau_k>0$.
\end{IEEEproof}

\begin{figure}[t]
\centering
\begin{tikzpicture}[x=0.118cm,y=0.118cm,>={Latex[length=1.6mm]},font=\scriptsize]
  \fill[pattern=north east lines,pattern color=black!45] (-3,15) rectangle (66,16.6);
  \draw[very thick] (-3,15)--(66,15);
  \draw[gray,dashed] (-3,18.4)--(66,18.4);
  \node[anchor=south west,inner sep=1pt] at (-3,18.6) {twin estimate of the wall};
  \node[anchor=north east,inner sep=1.5pt] at (66,15) {wall};
  \draw[black!35] (0,0)--(66,0);
  \draw[->,thick,targetblue] (0,0)--(59.2,0);
  \draw[->,thick,targetblue] (60,0)--(30,15);
  \draw[->,thick,targetblue] (30,15)--(1.2,0.6);
  \draw[->,thick,relayred,densely dotted] (30,0)--(15,15);
  \draw[->,thick,relayred,densely dotted] (15,15)--(1.0,1.0);
  \draw[targetblue,fill=white] (30,15) circle (0.8);
  \draw[relayred,fill=white] (15,15) circle (0.8);
  \draw[targetblue] (9,0) arc[start angle=0,end angle=26.57,radius=9];
  \node[targetblue] at (14.2,3.3) {$26.6^\circ$};
  \draw[relayred,densely dotted] (5.2,0) arc[start angle=0,end angle=45,radius=5.2];
  \node[relayred] at (6.6,8.6) {$45^\circ$};
  \fill[relayred] (30,0) circle (1.0); \fill[targetblue] (60,0) circle (1.0);
  \node[inner sep=0pt,anchor=north] at (0,0.752) {\includegraphics[height=0.944cm]{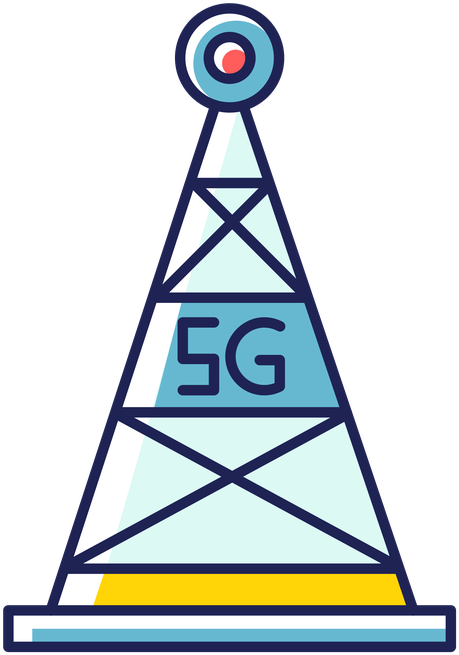}};
  \node[anchor=east] at (-3.2,-3.6) {gNB};
  \node[below] at (30,-1.2) {relay, $R_{\rm A}$};
  \node[below] at (60,-1.2) {claimed target, $R_{\rm T}$};
  \draw[thick,targetblue] (46,10.2)--(51,10.2); \node[anchor=west] at (51.3,10.2) {genuine echo};
  \draw[thick,relayred,densely dotted] (46,7.2)--(51,7.2); \node[anchor=west] at (51.3,7.2) {relay echo};
\end{tikzpicture}
\caption{Walls as mirrors. Solid: a genuine target's wall echo returns from $26.6^\circ$. Dotted: the relay's wall echo returns from $45^\circ$, whatever delay the relay adds.}
\label{fig:geometry}
\end{figure}

The relay's wall echo also comes from a different angle, the direction of $\mathcal M_k(\pA)$. On a street whose wall runs parallel to the bearing at lateral distance $d$, the echo angle is $\tan^{-1}(2d/R)$ and $u_k(R)=\sqrt{R^2+4d^2}-R$. For a claim at $R_{\rm T}=60$\,m, a relay at $R_{\rm A}=30$\,m, and $d=15$\,m, the genuine wall echo arrives from $26.6^\circ$ and the relay's from $45^\circ$, $17.8$\,ns later (Fig.~\ref{fig:geometry}). An error $\epsilon$ in the twin shifts the predicted echo by $2\cos\vartheta_k/c$ in delay and $2\sin\vartheta_k/\norm{\mathcal M_k(\mathbf p)-\pB}$ in angle per meter, with incidence angle $\vartheta_k$ to the wall normal ($63.4^\circ$ here), that is, by 3.0\,ns and $1.5^\circ$ per meter.

Could a cleverer relay shape its signal to fake the wall echo anyway? Not without giving itself away.

\begin{proposition}[Single-emitter coupling]
\label{prop:couple}
Consider a noiseless, clutter-free channel in which a single antenna at $\pA$ reaches the gNB through a LoS path and paths $i$ with at most two reflections. Let it transmit a signal $x\in\mathcal S$, the set of signals whose components reaching a delay window on the bearing through reflected paths also reach the reflector gate, the delay-angle window around the genuine wall echo, and whose LoS copies of those components land in the bearing window. On $\mathcal S$, let $g_{\rm L}$ be the smallest energy gain of the LoS path into the bearing window, and $e_i$ and $g_i$ the largest energy gains of path $i$ into the gate and the bearing window; the LoS path is assumed not to feed the gate (its angular sidelobes there are neglected). If the gate receives energy $E$, then the bearing window receives at least $\kappa_{\rm lb}E$, with
\begin{equation}
\kappa_{\rm lb}=\frac{\bigl(\sqrt{g_{\rm L}}-\sum_i\sqrt{g_i}\bigr)_+^2}{\bigl(\sum_i\sqrt{e_i}\bigr)^2}.
\label{eq:kappa}
\end{equation}
For energy reaching the gate through reflector $k$'s reciprocal path, the LoS copy arrives $\Delta\tau_k$ before the claimed direct echo, up to the gate width.
\end{proposition}

\begin{IEEEproof}
By the triangle inequality and the definition of $e_i$, the gate amplitude is at most $\sum_i\sqrt{e_i}\norm{x}$. The LoS path delivers at least $\sqrt{g_{\rm L}}\norm{x}$ to the bearing window, of which the reflected paths can cancel at most $\sum_i\sqrt{g_i}\norm{x}$; squaring and dividing gives~\eqref{eq:kappa}. The reciprocal path is $u_k(\pA)$ longer than the LoS path, so by~\eqref{eq:gap} its copy lies $\Delta\tau_k$ before the direct echo.
\end{IEEEproof}

To fake the wall echo the relay must also plant a second echo on the bearing, at least $\kappa_{\rm lb}$ times as strong, where a genuine target produces nothing. The relay can hide this echo only inside the real direct echo, which happens when $\Delta\tau_k$ is too small to resolve. The condition $x\in\mathcal S$ matters: a signal that reaches the bearing window only through the weaker reflected paths could partly cancel the forced echo, and bounding that case remains open. Clutter, deeper paths, and finite windows enter only through the certificate's margin. A relay is \emph{delay-resolved} when $\Delta\tau_k$ exceeds the \emph{point-spread width}, the $-3$\,dB width of the processed response to a single path (about $1.44/B$ with a Hann, or raised-cosine, taper). It is \emph{angle-resolved} when its wall echo arrives more than one beamwidth away from the genuine one. Both improve as the standoff $\Delta R=R_{\rm T}-R_{\rm A}$ grows.

\section{Twin-Guided Verification}
\label{sec:method}

The verifier turns this physics into a tool that the controller calls before acting on a candidate. It returns SUPPORTED (the wall evidence confirms a scatterer at the claimed position), UNRESOLVED, or ACQUISITION\_FAILED. A separate path status, NOT\_OBSERVABLE, records that the twin predicts no usable wall echo. No outcome ever declares an attack, so a missing echo never becomes an accusation.

\emph{Acquisition.} A first action beams toward the reported bearing. The receiver estimates the channel under beam $\mathbf w_b$ as $\widehat{\mathbf h}_b[n,m]=\mathbf y_b[n,m]s^*[n,m]$, applies Hann tapers across subcarriers and receive elements, and forms a zero-Doppler angle-delay map. The candidate is the strongest cell in a surveillance sector fixed before any data, provided it exceeds a floor $\gamma_{\rm acq}$; it yields $\widehat{\mathbf p}=\pB+(c\widehat\tau_0/2)\widehat{\mathbf u}$, with an uncertainty set by the point-spread width and the signal-to-noise ratio (SNR).

\emph{Planning.} The planner sees the twin, the candidate, the known symbols and beams, and $\sigma_{\rm map}$, but never the truth, the labels, or the relay. It places a hypothetical scatterer at $\widehat{\mathbf p}$ in a copy of the twin, lists the mixed paths with the image method, confirms them by ray tracing, and keeps the strongest one that departs at least $\phi_{\min}$ from the bearing; if none qualifies, the status is NOT\_OBSERVABLE.

\emph{Receive-side check.} The first action already contains the genuine target's reciprocal wall echo, so the twin simply says where to look in it; no extra transmission is needed. \emph{Wall probe.} Alternatively, one extra action lights the outgoing mixed path with $\mathbf w_{\rm W}\propto\mathbf P_\perp\overline{\mathbf v}_{\rm W}$, where $\mathbf P_\perp=\mathbf I-\mathbf v_0\mathbf v_0^{\rm H}/\norm{\mathbf v_0}^2$ and $\mathbf v_0$ points at the measured bearing, so the line of sight, where a relay must sit, is kept dark; the echo then returns along the bearing. The \emph{narrow} probe aims at the twin's nominal direction, and the \emph{robust} probe averages steering vectors over $N_{\rm s}$ samples of twin error and range uncertainty.

\emph{Evidence.} A \emph{gate} is a fixed delay-angle window whose energy is summed. The direct gate surrounds the candidate; the reflector gate surrounds the predicted wall echo, sized by the central 99\% of the twin samples, an inflation factor $c_{\rm g}$, the point-spread width, and the candidate uncertainty. From the gate energies $E_{\rm D}$ and $E_{\rm W}$, their calibrated noise levels $\widehat N_{\rm D}$ and $\widehat N_{\rm W}$, and the twin's predicted ratio $\widehat r_{\rm DT}$ under the same beams, the verifier forms
\begin{equation}
S=\frac{E_{\rm W}-\widehat N_{\rm W}}{\max(E_{\rm D}-\widehat N_{\rm D},\epsilon)\,\widehat r_{\rm DT}},\qquad \rho_{\rm W}=\frac{E_{\rm W}-\widehat N_{\rm W}}{\widehat N_{\rm W}},
\label{eq:score}
\end{equation}
with a small floor $\epsilon>0$. $E_{\rm D}$ comes from the first action, and $E_{\rm W}$ from the action carrying the wall echo: the first action for the receive-side check, and the probe, where the echo returns along the bearing, for the wall probe. A genuine target matching the twin scores $S\approx1$, and an unknown overall RCS cancels. The verdict is SUPPORTED when both echoes are detectable, $\rho_{\rm W}>k$, and $s_{\rm lo}\le S\le s_{\rm hi}$; the \emph{one-sided} rule omits $s_{\rm hi}$. The \emph{consistency check} applies Proposition~\ref{prop:couple}. It measures the energy on the bearing outside both gates, over a delay window that the twin derives from the paths a relay could use. It subtracts what the twin predicts there, and rejects the candidate when the remainder $C$ exceeds $c_{\rm hi}$.

\section{Evaluation Setup}
\label{sec:evaluation}

\emph{Scenario.} The gNB at $(0,0,2)$\,m has 16-element horizontal uniform linear arrays (ULAs) with half-wavelength spacing and operates at 3.5\,GHz. Each action carries 64 fresh quadrature phase-shift keying (QPSK) symbols of 18.75\,$\mu$s, including a 2.08\,$\mu$s cyclic prefix (1.2\,ms in total), with $\Delta f=60$\,kHz and $N=2048$, so $B=122.88$\,MHz; bandwidth stresses use $N=1024$ and 4096 with the same duration. The transmit power is $P_{\rm tx}=10$\,dBm and the noise figure $F=7$\,dB. Two vertical walls line a street at $y=\pm d$, modelled by Sionna RT as 0.10\,m slabs ($\epsilon_r=5$, 0.02\,S/m). The target is a static point with a constant 1\,m$^2$ RCS at gNB height, a stand-in for a hovering drone that ignores how real drone RCS varies with angle~\cite{semkin2020}. Target paths come from Sionna's RCS solver (depth 3) and clutter and relay links from its path solver (depth 2), depths at which one more bounce changes the power under the actual beams by less than 1\%. The fixed relay ($\tau_{\rm proc}=20$\,ns, $P_{\rm A}=10$\,dBm, $F=7$\,dB) is matched by the simulator; transmit-receive isolation is ideal. Acquisition searches $[300,500]$\,ns and $\pm60^\circ$ with $\gamma_{\rm acq}=10$\,dB, and planning uses $\phi_{\min}=10^\circ$ and $N_{\rm s}=257$.

\emph{Baselines.} Each baseline gets the same extra 64-symbol action; an exact-map oracle serves only as a reference. A coarse-to-fine search without a map sends four projected beams at $\pm30^\circ$ and $\pm50^\circ$ with 8 symbols each, then two beams $10^\circ$ either side of the strongest with 16 symbols each. It gates the 50\,ns after the measured direct echo at each beam angle, cannot form $S$, and is scored by $\rho_{\rm W}\ge t_m$ (SUPPORTED$_{\rm common}$). A registered random single-beam control was not analyzed.

\emph{Pre-registered protocol.} Every rule was committed to the code repository before the data it governs existed. Geometries draw $R_{\rm T}\sim U[50,70]$\,m, $R_{\rm A}\sim U[20,40]$\,m, and $d\sim U[12,18]$\,m: 12 for calibration and 24 held out. Each runs at $\sigma_{\rm map}\in\{0,0.25,0.5,1,2\}$\,m with 32 records per hypothesis, record $r$ using twin realization $r\bmod4$, for 7{,}680 held-out trials per method. Calibration data alone set the thresholds: $k$ allows 1\% noise-only exceedance; $s_{\rm lo}$ and $t_m$ are the smallest values that keep SUPPORTED under $\HA$ at or below 5\% for every $\sigma_{\rm map}$; $s_{\rm hi}$ is the 99.5th percentile of $S$ under $\HT$; $c_{\rm g}$ yields 99\% gate coverage; $c_{\rm hi}$ yields 1\% false rejection of genuine targets. The robust probe was the primary method, judged by the criteria of Table~\ref{tab:heldout}: CONDITIONAL GO if every point estimate meets its target, the replay rate's upper bound stays at or below 10\% in every cell, and the advantage's lower bound is positive; NO-GO if some whole 95\% interval lies on the failing side; INCONCLUSIVE otherwise. The geometry is the unit of independence, with 4 twin and 8 noise realizations per twin within it; intervals come from a 10{,}000-draw bootstrap over geometries, conditional on the frozen thresholds; for zero rates the one-sided 95\% Clopper-Pearson bound over 24 geometries (11.7\%) is also given. The receive-side check, the consistency check, and the controller came later: each was designed after the probe's held-out results and an external review, and registered before its own data. The receive-side check and the controller reuse the 24 held-out geometries. The certificate's window and margins were revised after their first evaluation, then frozen and re-tested on 24 fresh geometries drawn from a new seed stream. Two defects were fixed. An incoherent delay refinement was caught by the coverage check before any held-out data existed. A sign error in the receiver's angle axis was found during review; re-running two geometries with four records and all methods showed that its fix changed 3 of 480 decisions, all robust-probe decisions at the threshold.

\section{Results}
\label{sec:results}

\emph{Validation.} Over 70 beams in the setting of Proposition~\ref{prop:los}, ray-traced target and matched-relay channels differ by at most $8.3\times10^{-6}$. Ray-traced amplitudes match the radar equation and the image method to four digits, and the OFDM model matches a time-domain simulation to $2.1\times10^{-5}$. The point-spread width is 11.71\,ns in delay and $11.0^\circ$ in angle. For the probe, calibration gave $k=1.72$, $s_{\rm lo}=0.022$, $s_{\rm hi}=1.29$, and $c_{\rm g}=1$ with 100\% coverage.

\emph{The pre-registered verdict.} The probe confirmed genuine targets in 99.8\% of trials but also accepted 5.3\% of replays overall and 6.1\% at $\sigma_{\rm map}=0.5$\,m, just above the 5\% limit (Table~\ref{tab:heldout}), so the verdict is INCONCLUSIVE. Every false acceptance came from the 7 geometries whose $\Delta\tau_k$ falls below one point-spread width, as predicted, and the oracle failed on the same geometries (4.3\%). After the fact, the score itself told the two cases apart perfectly: the area under the receiver operating characteristic curve (AUC) is 1.000, and the largest $s_{\rm lo}$ keeping 99\% target support (0.65) would have accepted no replay. The miss thus came from a threshold set with almost no margin by a calibration set that held one hard geometry where the held-out set held seven, not from a lack of separability.

\begin{table}[t]
\caption{Pre-registered held-out criteria, robust probe, 24 geometries. Advantage: probe minus map-free target SUPPORTED rate at $\sigma_{\rm map}=0.5$ and $1\,\mathrm{m}$ (equal values).}
\label{tab:heldout}
\centering\footnotesize
\setlength{\aboverulesep}{0pt}\setlength{\belowrulesep}{0pt}\renewcommand{\arraystretch}{1.12}
\setlength{\tabcolsep}{2pt}
\begin{tabular}{@{}lccc@{}}
\toprule
\rowcolor{tabhead}Criterion & Target & Point & 95\% interval\\
\midrule
Target SUPPORTED, $\sigma_{\rm map}\le0.5\,\mathrm{m}$ & $\ge0.80$ & 0.998 & [0.996, 1.000]\\
\rowcolor{tabrow}Replay SUPPORTED, pooled & $\le0.05$ & 0.053 & [0.004, 0.131]\\
Replay SUPPORTED, $\sigma_{\rm map}=0.5\,\mathrm{m}$ & $\le0.05$ & 0.061 & [0.000, 0.158]\\
\rowcolor{tabrow}Replay SUPPORTED, $\sigma_{\rm map}=1\,\mathrm{m}$ & $\le0.05$ & 0.029 & [0.000, 0.074]\\
NOT\_OBSERVABLE & $\le0.20$ & 0.000 & 0/24 geometries\\
\rowcolor{tabrow}Advantage over map-free search & $>0$ & 0.132 & [0.029, 0.262]\\
\midrule
\rowcolor{tabhead}\multicolumn{4}{@{}l}{Verdict: INCONCLUSIVE}\\
\bottomrule
\end{tabular}
\end{table}

\emph{No extra transmission needed.} The receive-side check, calibrated by the same rules ($k=1.54$, $s_{\rm lo}=0.56$, $s_{\rm hi}=1.21$), confirmed every held-out genuine target and accepted no held-out replay. This was the opposite of the pre-registered prediction. Without thresholds, the two scores are nearly equivalent: AUC 1.000 on held-out data, at least 0.98 from 6\,m standoff upward, and 0.84 to 0.88 at 3\,m. At matched 99\% target support the probe accepts as few replays or fewer (11\% against 17\% at 6\,m); the gap under frozen thresholds reflects where each threshold landed. The evidence comes from the twin's prediction, not from the extra probe.

\begin{figure}[t]
\centering
\includegraphics[width=\columnwidth]{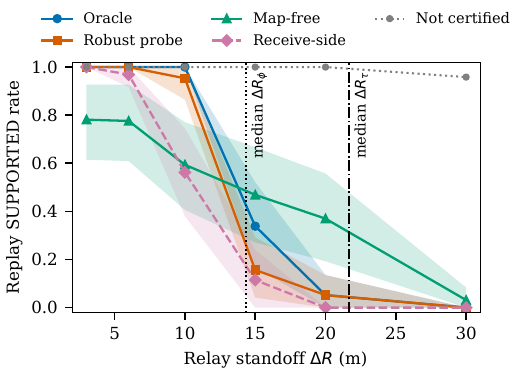}
\caption{Fixed-filter replays accepted versus relay standoff ($\sigma_{\rm map}=0.5$\,m, 24 held-out geometries, 95\% bands), and the fraction of geometries not certified by the conditional bound of Proposition~\ref{prop:couple} (grey). Vertical lines mark the median angular and delay standoffs, $\Delta R_\phi$ and $\Delta R_\tau$.}
\label{fig:standoff}
\end{figure}

\emph{How close can a relay hide?} Moving the relay $\Delta R=3$ to 30\,m in front of the claimed position (Fig.~\ref{fig:standoff}), the probe accepted 95\% to 100\% of replays up to 10\,m, 16\% at 15\,m, 5\% at 20\,m, and none at 30\,m. The receive-side check accepted 56\%, 12\%, 0\%, and 0\% at 10, 15, 20, and 30\,m, while the map-free search still accepted 47\% and 37\% at 15 and 20\,m. Acceptance falls near the median angular standoff (14.4\,m) rather than the delay standoff (21.7\,m), but because angle always resolved first here, the attribution is only suggestive. Doubling the bandwidth removed every false acceptance by the probe; halving it left the probe at 5.7\% but pushed the map-free method to 43\% in unresolved geometries.

\emph{A conditional certificate against single-antenna relays.} A geometry is certified when two conditions hold. First, the forced echo of every path, shifted over the whole reflector gate and widened by three standard deviations of twin and candidate error plus the point-spread width, lies in the consistency window and outside both gates. Second, $\kappa_{\rm lb}E_{\min}$, minus the energy the twin subtracts there, exceeds $c_{\rm hi}$ by 3\,dB, where $E_{\min}$ is the smallest gate energy that passes the receive-side thresholds. The certificate covers any signal that satisfies the conditions of Proposition~\ref{prop:couple}, sent from the evaluated relay position over the ray-traced channels; it is not yet a guarantee the verifier can issue for unknown relay positions. On these channels, $\kappa_{\rm lb}$ stays within 1.5\,dB of the single-path ratio, with a median of $+39$\,dB on held-out geometries. Its reach is small: 3 of 24 held-out geometries are certified, none at standoffs of 3 to 20\,m, and 1 of 24 at 30\,m. In every uncertified geometry, the widened forced echo overlaps the direct gate, so the certificate cannot separate it from the relay's own direct echo. This is a missing guarantee, not a demonstrated attack; the margins, not the bound, set the reach. On the 24 fresh geometries, the frozen design certified 3 of 24, none at 20\,m, and 2 of 24 at 30\,m. Bandwidth is the main lever (Table~\ref{tab:coverage}): on the same paths, the frozen certificate covers 0\%, 12.5\%, and 46\% of held-out geometries at 61.44, 122.88, and 245.76\,MHz for $\sigma_{\rm map}=0.5$\,m, and 63\% at 245.76\,MHz with a perfect twin, so twin error matters less.

\begin{table}[t]
\caption{Held-out geometries certified by the conditional bound (of 24), frozen design, same ray-traced paths.}
\label{tab:coverage}
\centering\footnotesize
\setlength{\aboverulesep}{0pt}\setlength{\belowrulesep}{0pt}\renewcommand{\arraystretch}{1.12}
\setlength{\tabcolsep}{4pt}
\begin{tabular}{@{}lccc@{}}
\toprule
\rowcolor{tabhead}$\sigma_{\rm map}$ & 61.44\,MHz & 122.88\,MHz & 245.76\,MHz\\
\midrule
0\,m & 0 & 4 & 15\\
\rowcolor{tabrow}0.25\,m & 0 & 4 & 14\\
0.5\,m & 0 & 3 & 11\\
\rowcolor{tabrow}1\,m & 0 & 3 & 11\\
\bottomrule
\end{tabular}
\end{table} The consistency threshold was set for 1\% false rejection on calibration data; it rejected 2.1\% $[0.1\%,4.4\%]$ of held-out genuine targets and 2.5\% on the fresh set, slightly above the pre-registered expectation of at most 2\%. Within these conditions the bound does not depend on the transmitted signal, so it also covers relays that adapt between transmissions.

\begin{figure}[t]
\centering
\includegraphics[width=\columnwidth]{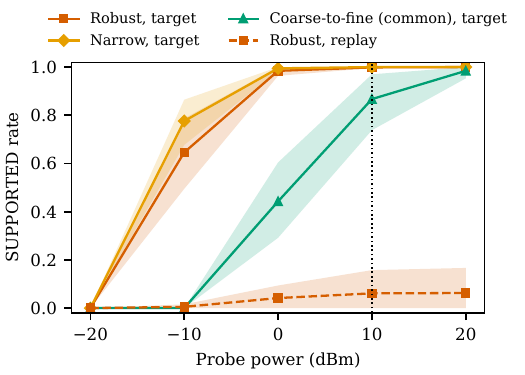}
\caption{Genuine targets confirmed versus probe power, and replays accepted by the robust probe (dashed); $\sigma_{\rm map}=0.5$\,m, 95\% bands.}
\label{fig:power}
\end{figure}

\emph{Power.} To confirm 80\% of genuine targets (interpolated on a 10\,dB grid), the probe needs $-5.4$\,dBm and accepts 2.2\% of replays, while the map-free search needs $+8.4$\,dBm and accepts 12.6\% (Fig.~\ref{fig:power}). The twin thus saves about 14\,dB of power while letting fewer replays through.

\emph{Robustness.} Table~\ref{tab:stress} lists the probe's stress tests. Array calibration errors of 0.5\,dB and $5^\circ$ root-mean-square (rms) per element made the null toward the relay shallower ($-34$\,dB instead of $-67$\,dB) without changing either rate. A thinner 0.05\,m wall raised replay acceptance to 13\%, so part of the margin comes from the wall material. An unknown 6\,dB rise in the target's reflected-path RCS made the two-sided rule reject every genuine target and the one-sided rule none, with unchanged replay rates; only $\pm6$\,dB was tested. The $-6$\,dB case passes only because the probe's $s_{\rm lo}$ is low. Fixed relay gain was not swept because $S$ is scale-invariant once $\rho_{\rm W}>k$.

\begin{table}[t]
\caption{Probe stress tests ($\sigma_{\rm map}=0.5\,\mathrm{m}$, frozen thresholds). Replay rates are identical under both rules.}
\label{tab:stress}
\centering\footnotesize
\setlength{\aboverulesep}{0pt}\setlength{\belowrulesep}{0pt}\renewcommand{\arraystretch}{1.12}
\setlength{\tabcolsep}{3pt}
\begin{tabular}{@{}lccc@{}}
\toprule
\rowcolor{tabhead}& \multicolumn{2}{c}{Target, rule} & \\
\rowcolor{tabhead}Level & two-sided & one-sided & Replay [95\% interval]\\
\midrule
Baseline & 0.999 & 1.000 & 0.061 [0.000, 0.158]\\
\rowcolor{tabrow}Bandwidth 245.76\,MHz & 0.995 & 1.000 & 0 of 24 geometries\\
Bandwidth 61.44\,MHz & 1.000 & 1.000 & 0.057 [0.000, 0.146]\\
\rowcolor{tabrow}Array calibration error & 0.995 & 1.000 & 0.057 [0.000, 0.146]\\
Power $-10$\,dBm & 0.646 & 0.740 & 0.005 [0.000, 0.016]\\
\rowcolor{tabrow}Wall 0.05\,m & 1.000 & 1.000 & 0.130 [0.031, 0.250]\\
Reflected-path RCS $-6$\,dB & 1.000 & 1.000 & 0.052 [0.000, 0.141]\\
\rowcolor{tabrow}Reflected-path RCS $+6$\,dB & 0.000 & 1.000 & 0.052 [0.000, 0.141]\\
\bottomrule
\end{tabular}
\end{table}

\emph{A closed decision loop.} A closed loop with fixed rules shows how a controller would use the verifier ($\sigma_{\rm map}=0.5$\,m). It runs the active probe as its first stage, which the receive-side check could replace to save one probe; if the result is SUPPORTED but $S$ is below $s_{\rm esc}=0.18$, the 1st percentile of genuine-target scores in calibration, it sends a fresh 245.76\,MHz probe and acts only if that one also confirms. On the held-out geometries it acted on none of 768 replays and withheld action from 1 of 768 genuine targets, escalating 47 times; a replay candidate cost 1.06 probes and 1.12 times the spectrum-time of one baseline probe, and a genuine target about one probe. At 15\,m standoff, with 8 records per geometry, it acted on 24 of 192 replays (12.5\%, $[1\%,26\%]$), at 1.14 probes and 1.28 spectrum-time per replay candidate, because the wider band cannot resolve those relays either.

\emph{Design implications.} Use the receive-side check by default and the probe for escalation; favor bandwidth over twin precision, since doubling the bandwidth nearly quadrupled certificate coverage while 1\,m twin errors barely mattered; and give thresholds margin by calibrating on geometries below the resolution limit.

\section{Limitations and Conclusion}
\label{sec:limits}

A relay that retunes between the first action and the probe defeats the probe. In the nominal geometry, fed through its wall path, it sets its internal delay to 182.3\,ns, 17.8\,ns less than in the first action, with about 47\,dB more gain and roughly $-22.6$\,dBm transmit power. The receive-side check with its consistency check withstands such relays only at certified relay positions and within the bound's conditions. It offers nothing against several cooperating relays, relays with antenna arrays, or a malicious reconfigurable surface that shapes its own reflections~\cite{shui2025}. The street is symmetric with two walls, and the target is a static point at gNB height; drone elevation, motion, propeller micro-Doppler, and angle-dependent RCS, relay hardware, self-interference, and communication users are not modelled. The twin errors studied are small offsets of known walls; an outdated twin with missing or new structures is a different and harder failure.

Walls let a twin predict where a genuine drone's echo must appear, and a lone relay cannot put its echo there without exposing itself. Against fixed-filter relays this evidence separates genuine from replayed candidates beyond about 10 to 20\,m, needs about 14\,dB less power than a search without a map, and can be read from the echo the base station already has. Against single-antenna relay signals that meet the bound's conditions, a certificate holds so far in only a few geometries, because uncertainty margins keep the forced echo from being separated from the direct echo; doubling the bandwidth raises this to 46\% of held-out geometries. The pre-registered probe verdict remains inconclusive because its threshold left no margin. This points to the next steps: larger calibration sets, margin-aware thresholds, elevated and moving targets, twins refined online from the same echoes with differentiable ray tracing~\cite{hoydis2023}, and a full cognitive controller, possibly an LLM agent, that calls this verifier before every consequential action.

\section*{Acknowledgment}
This material is based upon work supported in part by NSF (CNS-2120442, IIS-2325863) and NTIA (51-60-IF007). Any opinions, findings, and conclusions or recommendations expressed are those of the authors and do not necessarily reflect the views of NSF or NTIA. AI tools assisted with coding, text and figure editing; the authors verified all content.


\end{document}